\documentclass[pra,aps,twocolumn,showpacs,superscriptaddress]{revtex4-1}
\usepackage{verbatim}
\usepackage{amsmath}
\usepackage{graphicx}
\usepackage{amsfonts}
\usepackage{amssymb}
\usepackage{ulem}

\newcommand{\ipr}{Institut de Physique de Rennes, Universit\'e Rennes I - CNRS UMR 6251,Campus de Beaulieu, 35042 Rennes Cedex, France}
\newcommand{\ulb}{Universit\'e Libre de Bruxelles, Optique Non Lin\'eaire Th\'eorique, Campus Plaine, CP 231, 1050 Bruxelles, Belgium}

\begin{document}
\title{Excitable-like chaotic pulses in the bounded-phase regime of an opto-rf oscillator}
\author{M. Romanelli}
\email{marco.romanelli@univ-rennes1.fr}
\affiliation{\ipr}
\author{A. Thorette}
\affiliation{\ipr}
\author{M. Brunel}
\affiliation{\ipr}
\author{T. Erneux}
\affiliation{\ulb}
\author{M. Vallet}
\affiliation{\ipr}
\date{\today}

\begin{abstract}
We report theoretical and experimental evidence of chaotic pulses with excitable-like properties in an opto-radiofrequency oscillator based on a self-injected dual-frequency laser.  
The chaotic attractor involved in the dynamics produces pulses that, albeit chaotic, are quite regular: They all have similar amplitudes, and are almost periodic in time. Thanks to these features, the system displays properties that are similar to those of excitable systems. In particular, the pulses exhibit a threshold-like response, of well-defined amplitude, to perturbations, and it appears possible to define a refractory time. At variance with excitability in injected lasers, here the excitable-like pulses are not accompanied by phase slips.
\end{abstract}
\pacs{42.65.Sf, 42.60.Mi, 42.55.-f}
\maketitle

\section{Introduction}

Since its first appearence (in the Hodgkin-Huxley model of the ``squid giant axon''~\cite{Hodgkin1952}), excitability has proven to be a common feature of many disparate biological, chemical and physical systems, such as, for instance, the heart muscle, the neurons, the Belousov-Zhabotinsky reaction, liquid crystals, and so forth (see \cite{Epstein1998, Lindner2004, Izhikevich2007} and references therein). In optical systems, excitability has been reported for instance in optically trapped birefringent particles~\cite{Pedaci2011}, optoelectronics integrated circuits~\cite{Romeira2013}, or in laser systems, in presence of optical feedback~\cite{Giudici1997, Yacomotti1999}, of saturable absorbers~\cite{Dubbeldam1999, Selmi2014}, or under optical injection~\cite{Coullet1998,Wieczorek2002,Goulding2007, Kelleher2009, Kelleher2011, Turconi2013}.
The latter configuration, of particular interest for the present work, is conveniently described by Adler equation~\cite{Erneux2010} when the injected optical power is weak. In Adler model, excitability arises close to the saddle-node bifurcation marking the transition from phase-locking to phase unlocking. Therefore, the unmistakable experimental signature of an excitable pulse is a 2$\pi$ phase jump~\cite{Kelleher2009}. This is an example of an universal behavior because Adler equation is also well adapted to model coupled Josephson junctions~\cite{Wiesenfeld1996}, biological~\cite{Goldstein2009} or micromechanical~\cite{Agrawal2013} oscillators, and so forth.
In the present work, we consider an opto-RF oscillator based on a dual-frequency laser with frequency-shifted feedback. For low feedback levels, the phase of this system can also be described by Adler equation, and its dynamics is analogous to that of optically injected lasers~\cite{Thevenin2011a}. Another universal synchronization regime can be found when the coupling is not weak, and its effects on the amplitude of the oscillators cannot be neglected~\cite{Aronson1990}. In this case, the transition from phase-locking to phase-unlocking is not direct, but takes place through a window of frequency-locking without phase-locking, in which the relative phase is bounded~\cite{Thevenin2011}. This regime has some intriguing features~\cite{Romanelli2014, Thorette2016} and has attracted some interest recently, not only in the context of lasers~\cite{Kelleher2012, Ludge2012}, but also in hydrodynamics~\cite{Li2013} and in nanomechanical resonators~\cite{Barois2014}. It is thus natural to ask whether an excitable response is still possible in the bounded-phase case. 
In the present work, we provide theoretical and experimental evidence of a mechanism leading to an excitable-like response, occurring at the transition from the phase-locking to the bounded-phase regime. Our system bifurcates from a phase-locked to a chaotic, self-pulsating state in which the pulses are not accompanied by phase slips. In ``standard'' excitable systems, the self-pulsating state is associated to a simple attractor such as a limit cycle, and as a consequence the system always follows the same, unique path in phase space, and produces identical pulses. 
This property is not verified here. Nevertheless, albeit chaotic, the self-pulsating state is quite regular, because it consists of pulses of similar amplitudes, almost periodic in time (see inset of Fig.~\ref{fig:Fig2}(b) and Fig.~\ref{fig:exca} (b)).Thus, we found that the response associated to this particular chaotic attractor presents features that are similar to excitability: Existence of a threshold i.e. need of a \textit{finite} perturbation in order to trigger a response (which is fairly independent of the amplitude of the perturbation), and a well-defined refractory time during which the system cannot be excited again, after a first stimulus.
The paper is organized as follows. In the next section, we describe the model equations and present numerical bifurcation curves and diagrams. We show that a chaotic, bounded phase attractor exists, and that close to the bifurcation point, intensity pulses with excitable-like characteristics can be found. In section~\ref{sec:exp}, we describe the experimental setup and results, and compare them to the numerical predictions. Section~\ref{sec:con} is devoted to the conclusions. 
\section{Model and numerical results}\label{sec:theo}
\subsection{Model equations and bifurcation diagrams}\label{subsec:model}
We start our analysis from a rate--equation model that was introduced heuristically by Bielawski et al.~\cite{Bielawski1992} in order to study the dynamics of a two-polarization Nd--doped fiber laser. The model introduces a population inversion for each polarization mode. Physically, this comes from the fact that a given active ion will interact preferentially with a given polarization mode (depending on its orientation, which is perturbed by its local environment, and, in the case of a long active medium, on its position along the z--direction, because of longitudinal spatial hole burning~\cite{Zeghlache1995}). The two populations are coupled by the fact that an atom in the $x$ population can produce, by stimulated emission, a photon in the $x$ mode, but also, with a lower probability quantified by a coefficient $\beta$, a photon in the $y$ mode. We note that the presence of two population inversions can be justified theoretically on a more rigorous basis, starting from the Maxwell-Bloch equations~\cite{Zeghlache1995}. This model has also been derived starting from a Maxwell-Bloch approach by Chartier et al.~\citep{Chartier2000}. These rate equations permit to reproduce successfully the antiphase polarization dynamics of Nd-- and Er--doped fiber lasers~\cite{Bielawski1992,Lacot1994}, and also of bulk Nd:YAG lasers~\cite{Cabrera2005}. The antiphase oscillation frequency allows retrieving the value of the coupling coefficient $\beta$~\cite{Lacot1994}. 
The model equations, that wa have also used to analyze the synchronization dynamics of an opto-radiofrequency (opto-rf) oscillator, based on the beating between the two polarization modes of a self-injected, dual-frequency laser~\cite{Thevenin2012}, read as follows: 
\begin{eqnarray}
\frac{de_x}{ds} &=& \frac{\left( m_x + \beta m_y \right)}{1+\beta} \frac{e_x}{2}, \label{norm_e_x} \\
\frac{de_y}{ds} &=& \frac{\left( m_y + \beta m_x \right)}{1+\beta} \frac{e_y}{2} + i \Delta \, e_y + \Gamma e_x,\label{norm_e_y}\\
\frac{dm_{x,y}}{ds} &=& 1 - \left(|e_{x,y}|^2 + \beta |e_{y,x}|^2\right) \label{norm_m}\\ \nonumber
&-& \epsilon \, m_{x,y}\left[1 + (\eta - 1)(|e_{x,y}|^2 + \beta |e_{y,x}|^2)\right].
\end{eqnarray}
$e_{x,y}$ are the amplitudes of the two laser fields coupled by optical injection, and the relative population inversions $ m_{x,y}$.  
$\eta$ is the pump parameter, and $\epsilon$ is the inversion lifetime. 
\begin{figure}[htbp]
\centering
\includegraphics[width=\linewidth]{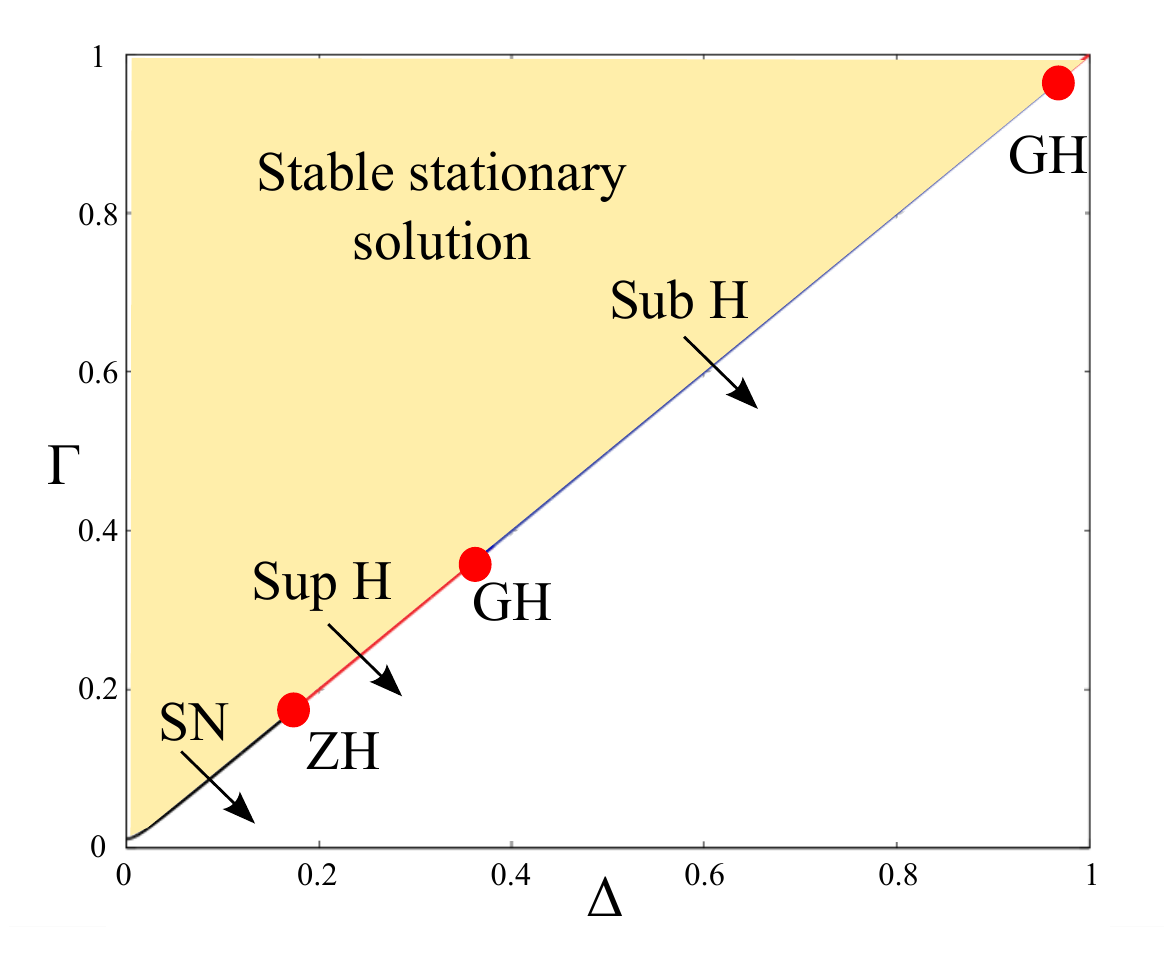}
\caption{Bifurcation curve. SN: saddle-node bifurcation. Sup (Sub) H: supercritical (subcritical) Hopf bifurcation. ZH and GH are codimension-two zero-Hopf and generalized-Hopf points respectively.}
\label{fig:bif_curve}
\end{figure}
The $e_x$ field is injected in the $e_y$ field. The injection process is described by two parameters, the detuning $\Delta$ and the injection strength $\Gamma$, which in the following are taken as the control parameters.  The other parameters are $\beta$ = 0.6, $\epsilon$ = 0.0097, and $\eta$ =1.2.
The scaled time $s$ is related to the physical time $t$ by $s= 2\pi f_R t$, where $f_R$ is the relaxation oscillation frequency. In our case $f_R \simeq 70$ kHz. Physically, the $e_y$ field is optically injected into $e_x$ via an external cavity (see Fig.~\ref{fig1}). The associated round-trip time $\tau_d \simeq 5$ ns is much smaller than $1/f_R$, so that the coupling can be considered instantaneous in the model. We have numerically checked the validity of this assumption.
In Fig.~\ref{fig:bif_curve}, we present a bifurcation curve in the parameter plane $\{\Delta, \Gamma \}$, computed using the continuation software MATCONT~\cite{Dhooge2003}. 
When $\vert \Delta \vert < \vert \Gamma \vert$, the model (\ref{norm_e_x}-\ref{norm_m}) admits a stable stationary solution, in which the laser fields are phase-locked. This solution bifurcates to a time-dependent state when $\vert \Delta \vert$ becomes larger than $\vert \Gamma \vert$. According to~\cite{Izhikevich2000}, excitability arises when a system is ``near a bifurcation (transition) from quiescence to repetitive firing''. Therefore in our system excitability has to be searched close to the boundary of the phase-locking range.
\begin{figure}[htbp]
\centering
\includegraphics[width=\linewidth]{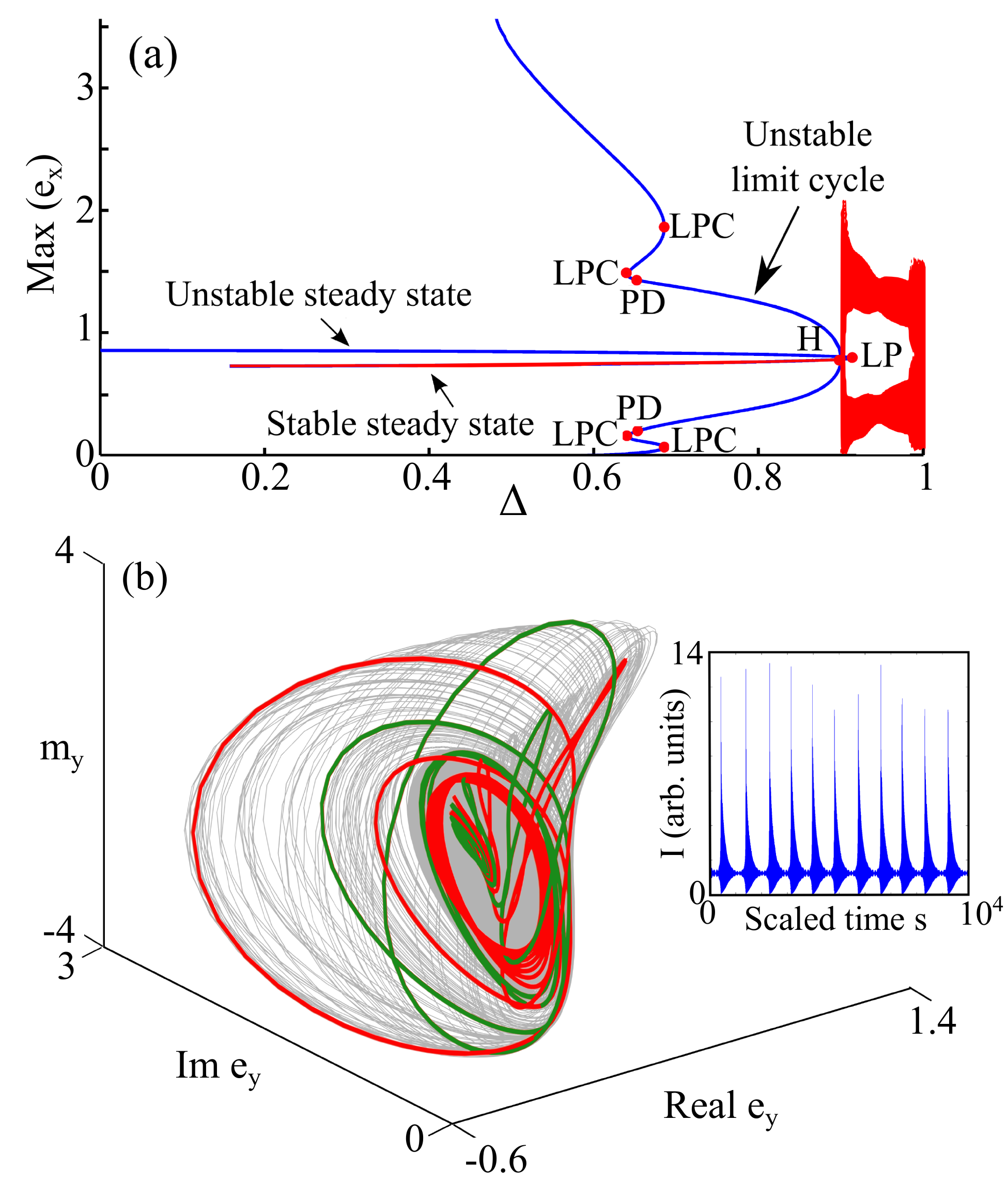}
\caption{(a) Blue: bifurcation diagram as a function of the control parameter $\Delta$, calculated using continuation methods. Red: bifurcation diagram calculated by numerical integration of eqs.~(\ref{norm_e_x}-\ref{norm_m}). The bifurcation points are: H, Hopf; PD, period doubling; LP, limit point (saddle-node bifurcation); LPC, limit point of cycles (saddle-node bifurcation of limit cycles).
(b) Projection of the dynamics in a tridimensional phase space. The attractor is reconstructed using a time series calculated with a standard 4$^{th}$ order Runge-Kutta method. The integration step is 0.1, and the time series length is 50 000 in normalized units. The whole time series contains about 50 spikes. Inset: a part of the intensity time series. $\Gamma = 0.9$, and $\Delta = \Gamma + 10^{-3}$. The green and red trajectories on the attractor correspond to two different spikes.}
\label{fig:Fig2}
\end{figure}
Fig.\ref{fig:bif_curve} shows that the nature of the bifurcation to the time-dependent state depends on the values of $\Delta$ and $\Gamma$. For low values of $\Delta$ and $\Gamma$, the stationary phase-locked state is destabilized by a saddle-node (SN) bifurcation, leading directly to a phase-unlocked regime. In this range of parameters, the Adler approximation applies. For higher values of the parameters, the SN point becomes a Hopf (H) bifurcation point. In this case, the time-dependent state arising after the bifurcation corresponds to the bounded-phase regime, in which, contrary to the Adler case, synchronization is preserved~\cite{Romanelli2014}. The transition from SN to H bifurcation occurs at the codimension-two zero-Hopf point~\cite{Kuznetsov1998} labelled ZH in Fig.~\ref{fig:bif_curve}.
Furthermore, the Hopf bifurcation can be either supercritical or, for $0.36 < \Delta \simeq \Gamma < 0.97$, subcritical.
This region is of particular interest for the present work, because a subcritical bifurcation causes the system to jump to a distant region in phase space, and thus tends to promote a response involving large pulses. 
We underline that this subcritical bifurcation does not exist if the two laser modes are not coupled by the cross saturation $\beta$ in the active medium. Indeed, for $\beta$ = 0 the $e_x$ field and the respective population are decoupled from the other variables, and the model reduces to the description of an optically injected laser, for which the bifurcation is always supercritical~\cite{Thevenin2012}. In this respect, when the feedback is not weak the system we study here differs fundamentally from standard optical injection. 

We now focus on the region containing the subcritical bifurcation, and analyze a bifurcation diagram computed using $\Delta$ as a control parameter, for a fixed value of $\Gamma$, that we take equal to 0.9 in the following. 
A rather complex bifurcation structure, shown in Fig.~\ref{fig:Fig2} (a), is uncovered. The phase-locked state is destabilized by the subcritical Hopf bifurcation, where an unstable limit cycle appears. This cycle can be tracked by continuation, and it undergoes several secondary bifurcations without folding back towards the $\Delta > \Gamma$ region (Fig.\ref{fig:Fig2} (a), blue curve). Indeed, numerical integration of the model equations (\ref{norm_e_x}-\ref{norm_m}) shows that, when $\Delta$ becomes larger than $\Gamma$, a direct transition from the stationary state to deterministic chaos occurs~\cite{Thorette2016} (Fig.~\ref{fig:Fig2} (a), red curve). Fig.~\ref{fig:Fig2} (b) shows a projection of the chaotic attractor in the \{Real($e_y$), Im($e_y$),$m_y$\} space, together with a time series of the beat-note intensity $I = \vert e_x+e_y \vert^2$, consisting of a train of spikes that are quite regularly spaced. Two parts of a trajectory on the attractor are shown in red and green, corresponding to two different intensity spikes. It can be seen that, even if the intensity associated to the two spikes is very close, they actually correspond to well separated paths in phase space. The chaotic nature of the attractor has been assessed by calculating its largest Lyapunov exponent. 
To summarize, the bifurcation from quiescence to repetitive firing occurs at $\Delta \simeq \Gamma$, from a stable steady-state (phase-locking) to a self-pulsating chaotic state.  
\subsection{Excitable-like properties of the chaotic pulses}
\begin{figure}[htbp]
\centering
\includegraphics[width=\linewidth]{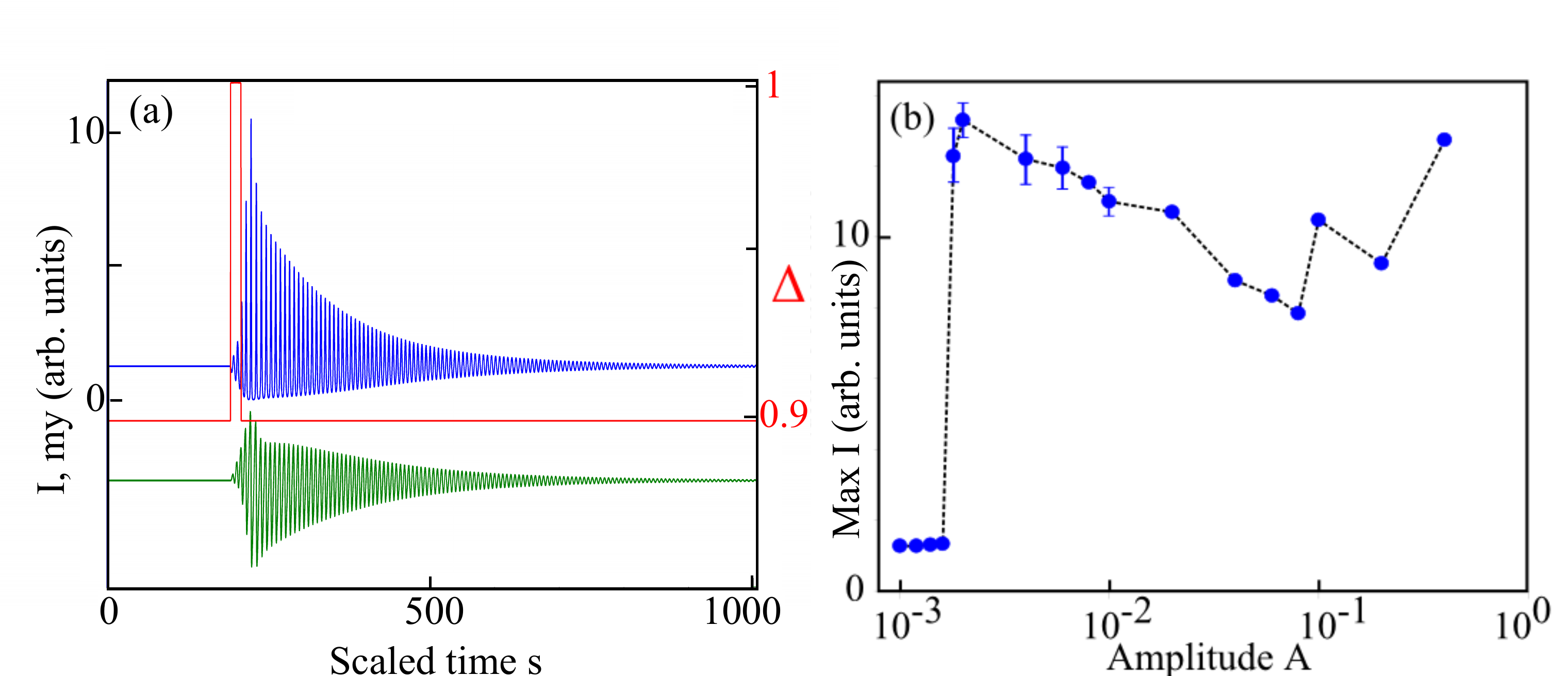}
\caption{(a) Response to a single perturbation of the detuning $\Delta$. Upper (blue) trace: beat-note intensity. Lower (green) trace: population inversion. (b) Amplitude of the response as a function of the perturbation amplitude $A$.}
\label{fig:exca}
\end{figure}

\begin{figure}[htbp]
\centering
\includegraphics[width=\linewidth]{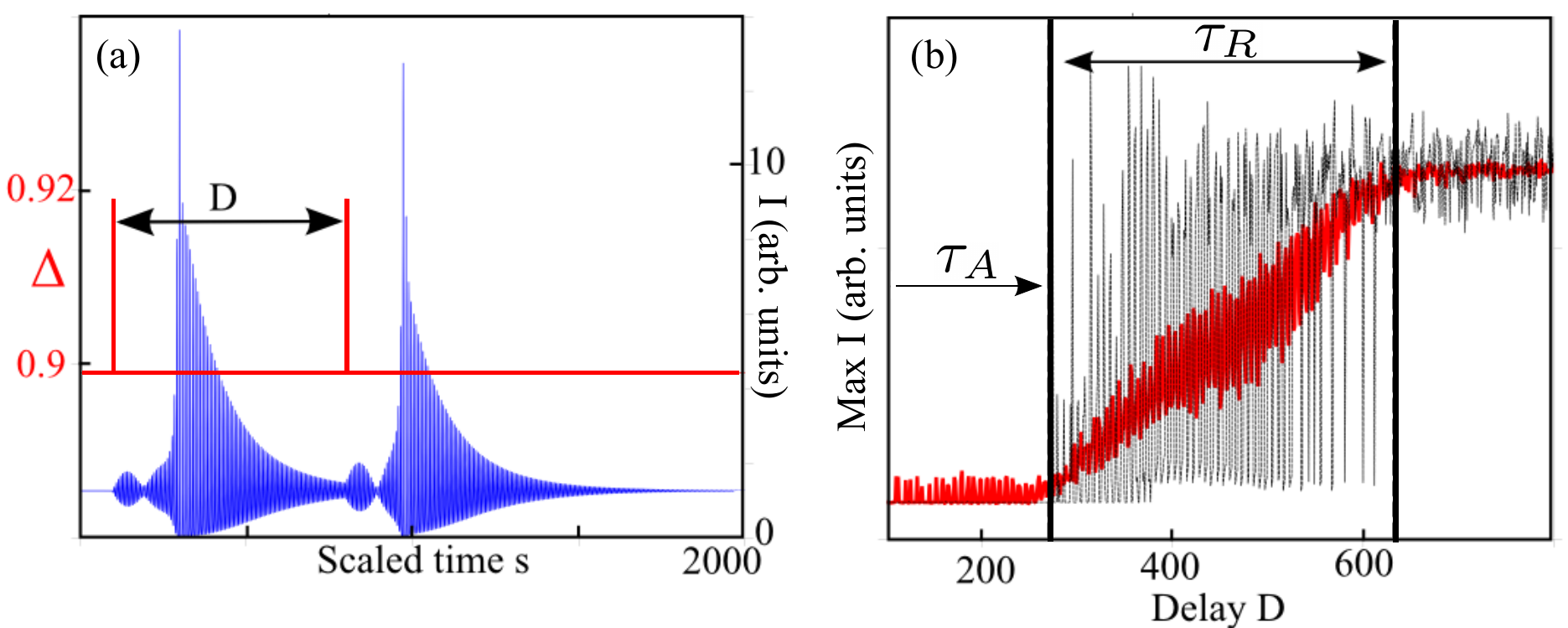}
\caption{(a) Response to two perturbations separated by a delay $D$. (b) Amplitude of the response to the second perturbation as a function of $D$. Dashed black curve: single realization. Red line: average over 40 realizations. $\tau_A$: absolute refractory period. $\tau_R$: relative refractory period.}
\label{fig:excb}
\end{figure}

The response of a system to a perturbation is called excitable if it presents the following features~\cite{Krauskopf2003}. First, it must have a \textit{all-or-none} character. This means that perturbations do not trigger a response if their amplitude is below a certain threshold, while perturbations of sufficient amplitude trigger a response which consists in a large excursion away from equilibrium, and does not depend on the amplitude of the perturbation. Furthermore, excitable systems possess a well defined refractory time during which they cannot be excited again, after a first stimulus.
In standard excitable systems, the system is in a quiescent state close to a simple limit-cycle attractor, induced by a Hopf or a homoclinic bifurcation leading to regular oscillations or pulses. In our case, there is a chaotic attractor, and the pulses do not follow an unique path in phase space. Despite this important difference, the system's response to perturbations of the detuning parameter $\Delta$ still exhibits features that are reminiscent of those of excitable systems (Figs.~\ref{fig:exca}-~\ref{fig:excb}).
\par First, we have checked that the response to a perturbation has a well-defined threshold, and is fairly independent of the perturbation's amplitude when the threshold is exceeded. 
In order to study the response to a single perturbation, a deterministic ``kick'' (whose analytical form is a very steep supergaussian function $p(x) = A\exp(-x^n)$, with $x= \frac{s-s_0}{W/2}$ and $n=1000$, W being the duration of the kick) was added to $\Delta$ at a given instant $s_0$, and its effect on the beat-note intensity $I=|e_x+e_y|^2$ was computed (Fig.~\ref{fig:exca}). Figs.~\ref{fig:exca} (a) and Fig.~\ref{fig:excb}(a) show that the perturbation excites the relaxation oscillations. A large pulse is produced, followed by oscillations at the relaxation oscillation frequency of both the intensity and the population inversion.
To produce the Fig.~\ref{fig:exca} (b), we have plotted the maximum of $I$ as a function of the perturbation amplitude $A$, the perturbation ``energy'' $A W$ being held constant at the value of 1.6.
For each point, we have integrated the equations 100 times, each time taking a different initial condition (the final point of the previous integration, i.e. a point which is very close to the fixed point corresponding to the locked state). This should mimick a real experiment repeated sequentially 100 times. The duration of each time series was 3 000 in normalized units. For a given value of $A$, some variation in the response (illustrated by the error bars, visible only for some points of Fig.~\ref{fig:exca}(b)) can be observed; since there is no noise in the system, this variation is due to sensitivity to the initial conditions and is a signature of the chaotic nature of the attractor. However, it is clear that this variation is relatively small, and that, despite the presence of chaos, a response of well-defined amplitude can be associated to a given perturbation.
The amplitude of the response shows some dependence on the amplitude of the perturbation, varying from around 7.8 to 13.3 while $A$ changes from $ 10^{-3}$ to $0.4$.
\par Second, we have found that, after a first excitation, well-defined time intervals during which the system's response is either completely or partially inihibited can be identified (Fig.~\ref{fig:excb}). These time intervals are reminiscent of the absolute and relative refractory times of excitable systems, as we will discuss below.
In order to investigate this property, we have submitted the system to two "kicks", with variable delay $D$ between them (see Fig.~\ref{fig:excb}(a)).
Again, the crucial point is the repeatability of the response for a fixed delay $D$. So, it is mandatory to repeat the same sequence of two pulses many times, starting from different initial conditions. This is how the reponse curve in Fig.~\ref{fig:excb}(b) has been obtained. 
In order to mimick a real experiment, we have taken, for each given value of the delay $D$, a sequence of 40 double pulses. Two consecutive double excitations are separated by a time interval of 2 000 in normalized units. The response as a function of $D$, averaged over 40 double excitations, is displayed in Fig~\ref{fig:excb}(b).
An \textit{absolute} refractory period $\tau_A \simeq 280$ can be unambiguously identified. If $D < \tau_A$, then the system can never be triggered by the second perturbation. The absolute refractory period $\tau_A$ can be understood as follows. Looking at Fig.~\ref{fig:excb}(a), it can be observed that the intensity spike appears with a substantial time-delay after the kick. It is this time-delay which determines $\tau_A$: If the second kick arrives before that the intensity spike has developed, then it will not generate a response. The time-delay depends mainly on the distance from the bifurcation point after the perturbation, i.e. on the initial value of the detuning $\Delta$ and on the size of the perturbation, as can be observed by comparing the responses in Fig.~\ref{fig:exca}(a) and Fig.~\ref{fig:excb}(a). In particular, a larger perturbation determines a faster response. So, also $\tau_A$ depends on the distance from the bifurcation point, i.e. on the detuning $\Delta$ and on the size of the perturbation. On the contrary, it does not depend much on which path the system follows in phase space: This makes it possible to observe even in our case a feature similar to the refractory time of excitable systems.
\par If the delay $D$ between the two kicks is larger than $620$, a second excitation always triggers a response.
In excitable systems, there exists also a \textit{relative} refractory period $\tau_R$, during which the response of the system is weaker, but not completely inhibited~\cite{Selmi2014}.
Interestingly, the delay interval $280 < D < 620$ can be identified, in a sense which will be precised in a moment, with the relative refractory period.
The dashed black curve in Fig.~\ref{fig:excb}(b) represents the response to a single double perturbation, as a function of $D$. It can be seen that, in the relative refractory period, the system's response depends on $D$ in a seemingly random fashion. Indeed, we have found that, for a given delay inside the relative refractory period $\tau_R$, the response is not always the same for two consecutive double excitations, i.e. for different initial conditions. Since our model is deterministic, this is, again, an effect of the sensitivity of chaotic systems to the initial conditions. However, it can be seen that, when averaging over several realizations, the \textit{probability} of triggering a second pulse increases linearly from 0 to 1 when $D$ goes from 280 to 620, thus producing a response curve which is very similar to the one presented, for instance, in Fig. 3(b) of ~\cite{Selmi2014}, where the relative refractory period of an excitable semiconductor laser is investigated.
In this sense, the [280,620] interval is reminiscent of the relative refractory period. We stress that the persistence of these features, typical of excitability, even in a chaotic situation, depends on the particular nature of the chaotic attractor we deal with. First, the pulses, albeit chaotic, all have very similar amplitudes, so that a well-defined response is obtained. Second, even if the pulses are chaotic, and are thus associated with different trajectories in phase space, a well-defined refractory time can still be defined, because it is essentially determined by the distance from the bifurcation point.
Thanks to these characteristics, the system's behavior is effectively close to an excitable one. Of course, in general this is not expected to be the case when chaos is present~\cite{Zamora2013}. Loosely speaking, it can be expected that a behavior similar to excitability may be obtained only if, as in our case, the chaotic self-pulsating state does not differ too much from a perfectly periodic one. 
\section{Experimental results}\label{sec:exp}
\subsection{Experimental setup}
\begin{figure}[hbp]
\centering\includegraphics[width=8cm]{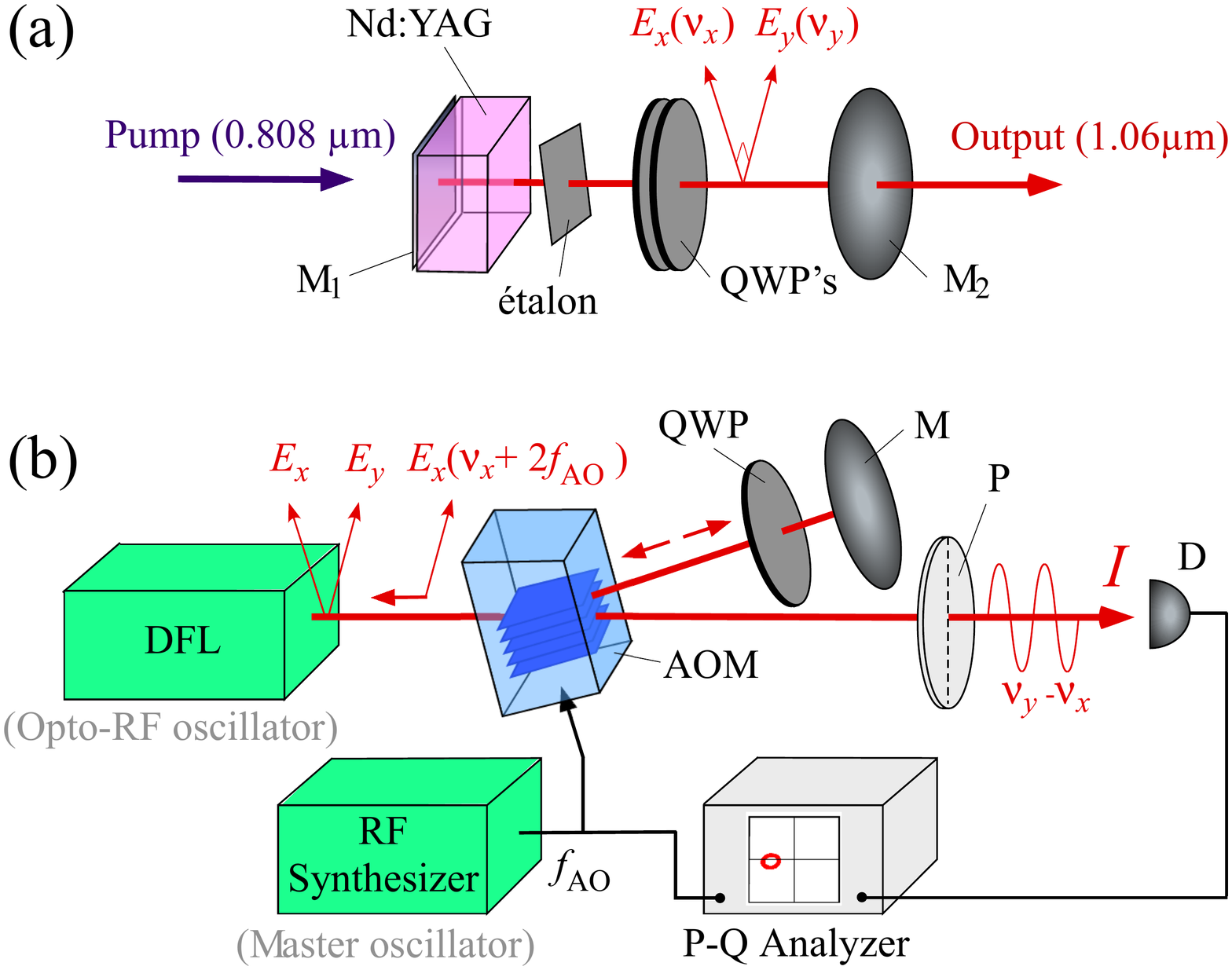} \caption{(a) Dual-Frequency Laser. $M_{1,2}$: cavity mirrors. QWP: quarter-wave plate. (b) Experimental setup that allows synchronizing the DFL beat-note frequency to a RF synthesizer. DFL: Dual-frequency laser (slave oscillator). M: feedback mirror. QWP: quarter-wave plate. AOM: acousto-optic modulator. P: polarizer. D: detector. P-Q analyzer: digital vector signal analyzer, permitting to measure the quadratures of \textit{I} with respect to the reference signal delivered by the RF synthesizer.} \label{fig1}
\end{figure}
The numerical results can be compared to experiments using the opto-radiofrequency oscillator described for instance in~\cite{Romanelli2014}. This system produces an optically-carried radiofrequency signal, thanks to the interference between the two modes of a dual-frequency laser (see Fig.~\ref{fig1}). 
The experimental setup is as follows. The laser cavity, of length L = 75 mm, is closed on one side by a high-reflection plane mirror, coated on the 5-mm long Nd:YAG active medium, and on the other side by a concave mirror (radius of curvature of 100 mm, intensity transmission of 1\% at the lasing wavelength $\lambda$ = 1064 nm). The active medium is pumped by a laser diode emitting at 808 nm. A 1 mm-thick silica \'etalon ensures single longitudinal mode oscillation. Two eigenmodes $E_x$ and $E_y$, polarized along $\hat{x}$ and $\hat{y}$, with eigenfrequencies $\nu_x$ and $\nu_y$ respectively, oscillate simultaneously. An intracavity birefringent element (here two quarter-wave plates QWPs) induces a frequency difference, finely tunable from 0 to $\frac{c}{4L}$ = 1 GHz by rotating one QWP with respect to the other~\cite{Brunel1997}. Here $\nu_y -\nu_x \simeq 180 $ MHz. The typical output power of the two-frequency laser is 10 mW when pumped with 500 mW.
When the laser output is detected by a photodiode after a polarizer at 45$^\circ$, an electrical signal oscillating at the frequency difference $\Delta \nu_0= \nu_y - \nu_x$ is obtained. The DFL can thus be seen as an opto-RF oscillator. In order to lock this oscillator to an external reference signal, we use optical frequency-shifted feedback~\cite{Kervevan} (Fig.~\ref{fig1}(b)). The feedback cavity contains an acousto-optic modulator (AOM), driven by a stable RF synthesizer, which provides an external phase reference. Next, a quarter-wave plate at 45$^\circ$ followed by a mirror flips the $\hat{x}$ and $\hat{y}$ polarizations, and finally the laser beam is reinjected in the laser cavity after crossing again the AOM. As a result, a $\hat{x}$-polarized field oscillating at the frequency $\nu_y+2 f_{AO}$ and a $\hat{y}$-polarized field oscillating at the frequency $\nu_x+2 f_{AO}$ are reinjected in the laser. We choose the value of $2 f_{AO}$ so that $\nu_x+2 f_{AO}$ is close to $\nu_y$. Under suitable feedback conditions, $\nu_y$ locks to the injected beam frequency $\nu_x+2 f_{AO}$, i.e. the frequency difference $\nu_y -\nu_x$ locks to $2 f_{AO}$. We note that the optical reinjection has no direct effect on $E_x$, because the frequency difference between $\nu_x$ and $\nu_y+2 f_{AO}$ is too large. For the same reason, multiple round trips in the feedback cavity have no effect on the dynamics. The laser output is detected with a photodiode (3 GHz analog bandwidth) after a crossed polarizer, thus providing an electrical signal proportional to $I=|E_x+E_y|^2$. The signal is then analyzed with an electrical spectrum analyzer, a digital P-Q signal analyzer and an oscilloscope.
\subsection{Bounded-phase pulses}
We call Adler frequency $f_A$ the maximum value of the detuning $\Delta \nu$ for which phase-locking can be achieved. In the experiments, the feedback strength is set in order to have $f_A$ a little smaller than the relaxation oscillation frequency $f_R \simeq$ 70 kHz. Typically $f_A = 0.9 f_R$ i.e. $\Gamma = 0.9 $.  The detuning is set in order to put the opto-rf oscillator very close to the boundary of the phase-locking range, i.e. to the Hopf bifurcation point: $\Delta \nu \lesssim f_A$. Typically $f_A-\Delta \nu \simeq 100 $ Hz, i.e. $|\Delta-\Gamma| \ll 1$. 
\begin{figure}[htbp]
\centering
\includegraphics[width=\linewidth]{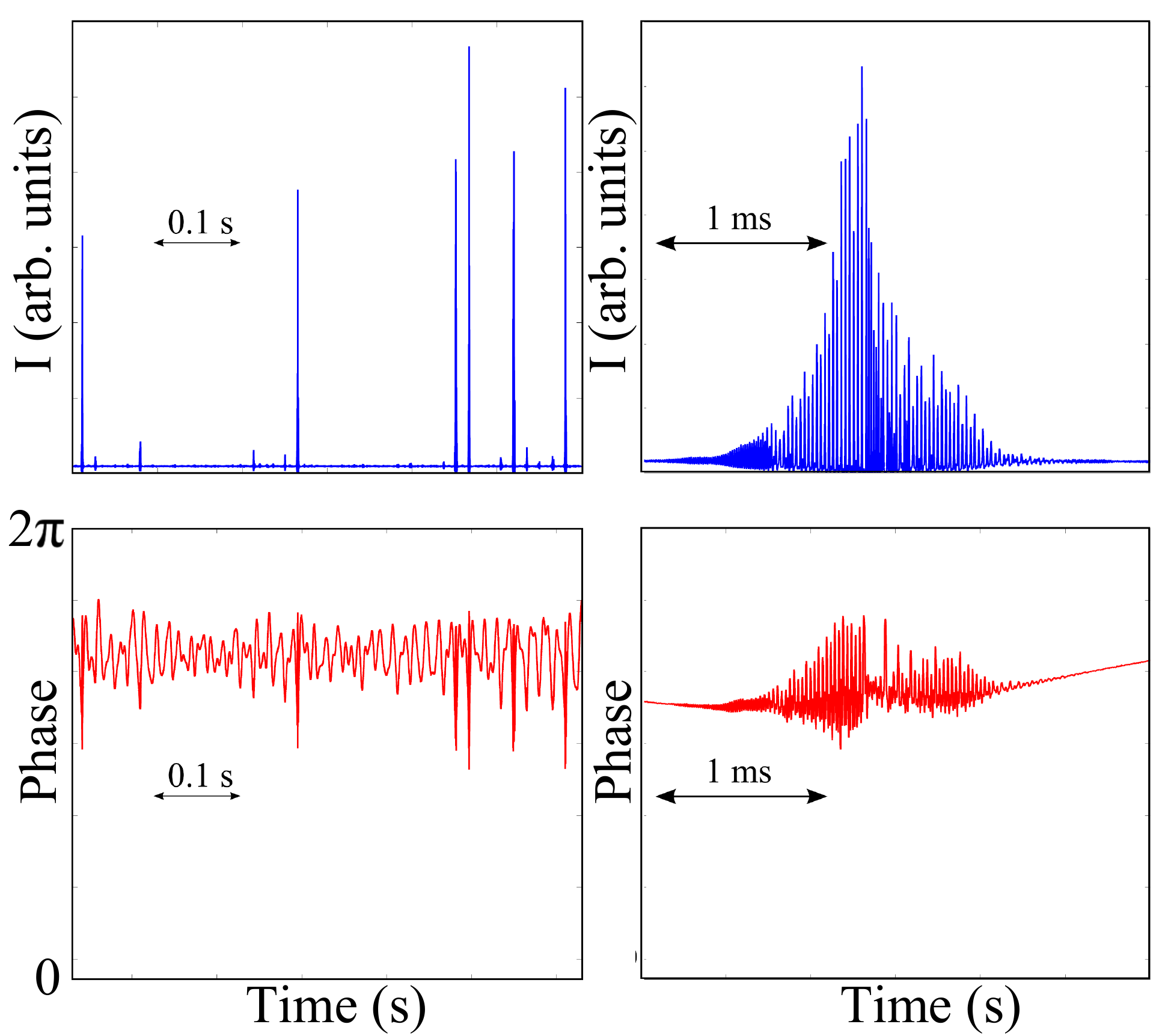}
\caption{Experimental intensity (top) and phase (bottom) time series. Note that in this figure the DC part of the signal has been removed for technical reasons.}
\label{fig:Fig3}
\end{figure}
An experimental time series of the beat-note intensity and phase under these experimental condition is shown in Fig.~\ref{fig:Fig3}.
Large spiking events appear in the time series at random instants. When observed at a shorter time scale, each isolated event consists of a bunch of pulses, originated by strong oscillations at the relaxation oscillation frequency. From the experimental phase time series, it is clearly seen that the phase of the beat-note signal is weakly affected by an intensity spike. In particular, the phase remains bounded throughout all the bunch of pulses. This differs strongly from previous reports on excitability generated by saddle-node bifurcations and explained by Adler mechanism~\cite{Goulding2007, Kelleher2009, Turconi2013}, and also from multipulse excitability~\cite{Wieczorek2002, Kelleher2011}, where an excitable pulse is necessarily accompanied by a 2$\pi$ phase jump, as experimentally demonstrated in~\cite{Kelleher2009}. 
\begin{figure}[floatfix]
\centering
\includegraphics[width=\linewidth]{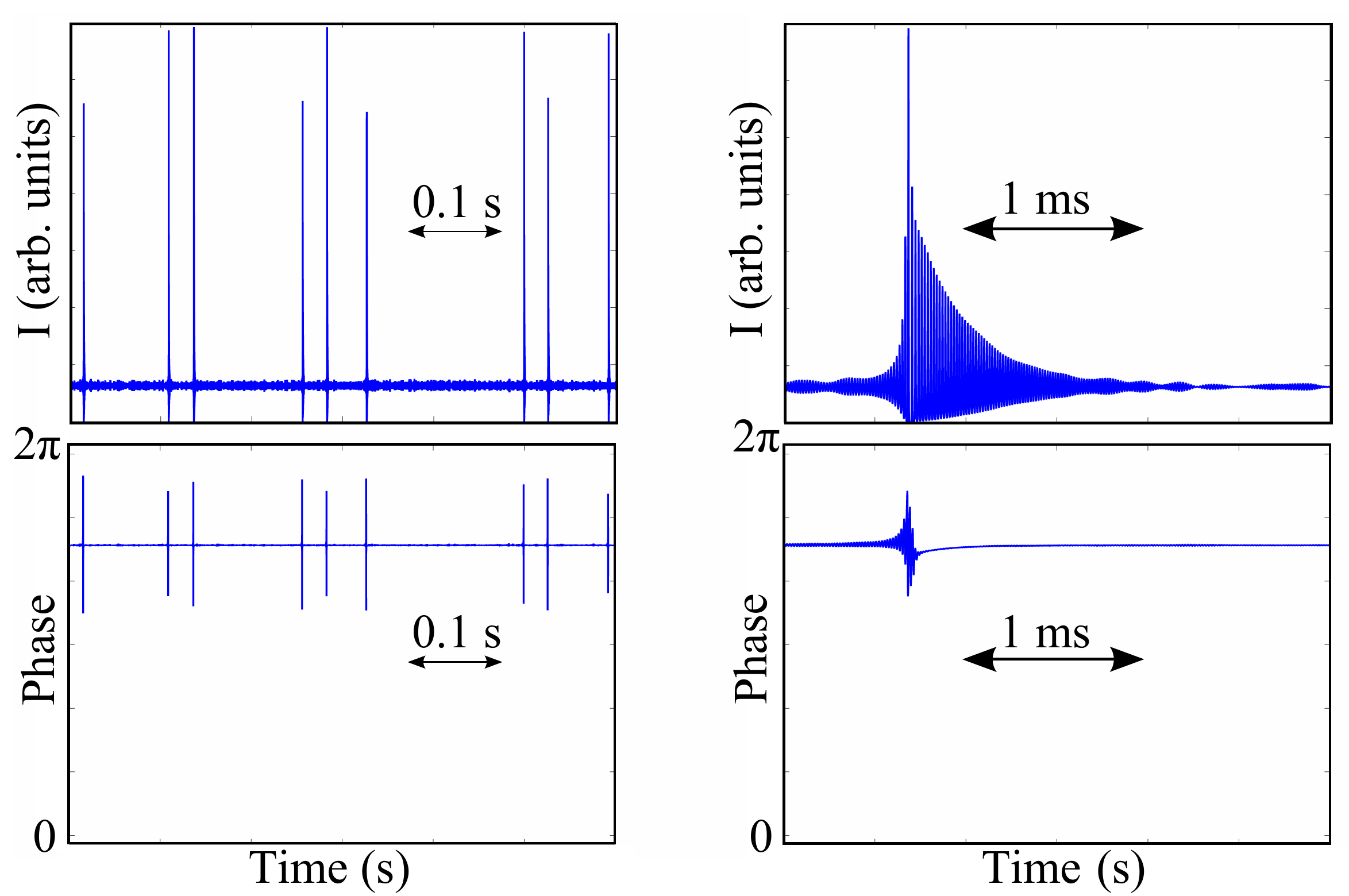}
\caption{Computed beat-note intensity (top) and phase (bottom) time series, using equations (\ref{norm_e_x}-\ref{norm_m}). The parameter values are $\beta$ = 0.6, $f_R$ = 70 kHz, $\epsilon$ = 0.0097, $\eta$ = 1.2, $\Gamma$ = 0.9, and $\Delta$ = $\Gamma - 10^{-3} + \xi(s)$, where $\xi(s)$ is a normally distributed stochastic process with a standard deviation $\sigma = 5 \; 10^{-3}$.}
\label{fig:Fig1}
\end{figure}
By comparing the experimental intensity time series and the inset of Fig.~\ref{fig:Fig2} (b), it appears clearly that frequency noise has to be included in the model in order to reproduce the experimental observations. Indeed, the time interval between two chaotic spikes is very regular in a purely deterministic model, in evident contrast with the experiments. So, we interpret the experimental observations as follows: The oscillator is actually inside the phase-locking range, but very close to the bifurcation point, so that we observe noise-induced pulses as in~\cite{Kelleher2009}. A simulation taking into account these observations is presented in Fig.~\ref{fig:Fig1}, which shows a calculated time series of the beat-note intensity $I = \vert e_x+e_y \vert^2$ and of the relative phase. In this simulation, the system is close to the boundary of the phase-locking range, and the detuning parameter includes an additive stochastic contribution $\xi(s)$ in order to account for the experimentally observed fluctuations of the beat-note frequency when the laser is free running. These fluctuations are typically of the order of some tens of Hz on a 1 s time scale. A good qualitative agreement with experiments is found. In particular it is confirmed that the phase of the beat-note signal is weakly affected by an intensity spike. The size of the phase excursion during the spike appears very close to the experimental observations.
\subsection{Experimental tests of excitable-like properties: all-or-none response, refractory period}
\begin{figure}[htbp]
\centering
\includegraphics[width=\linewidth]{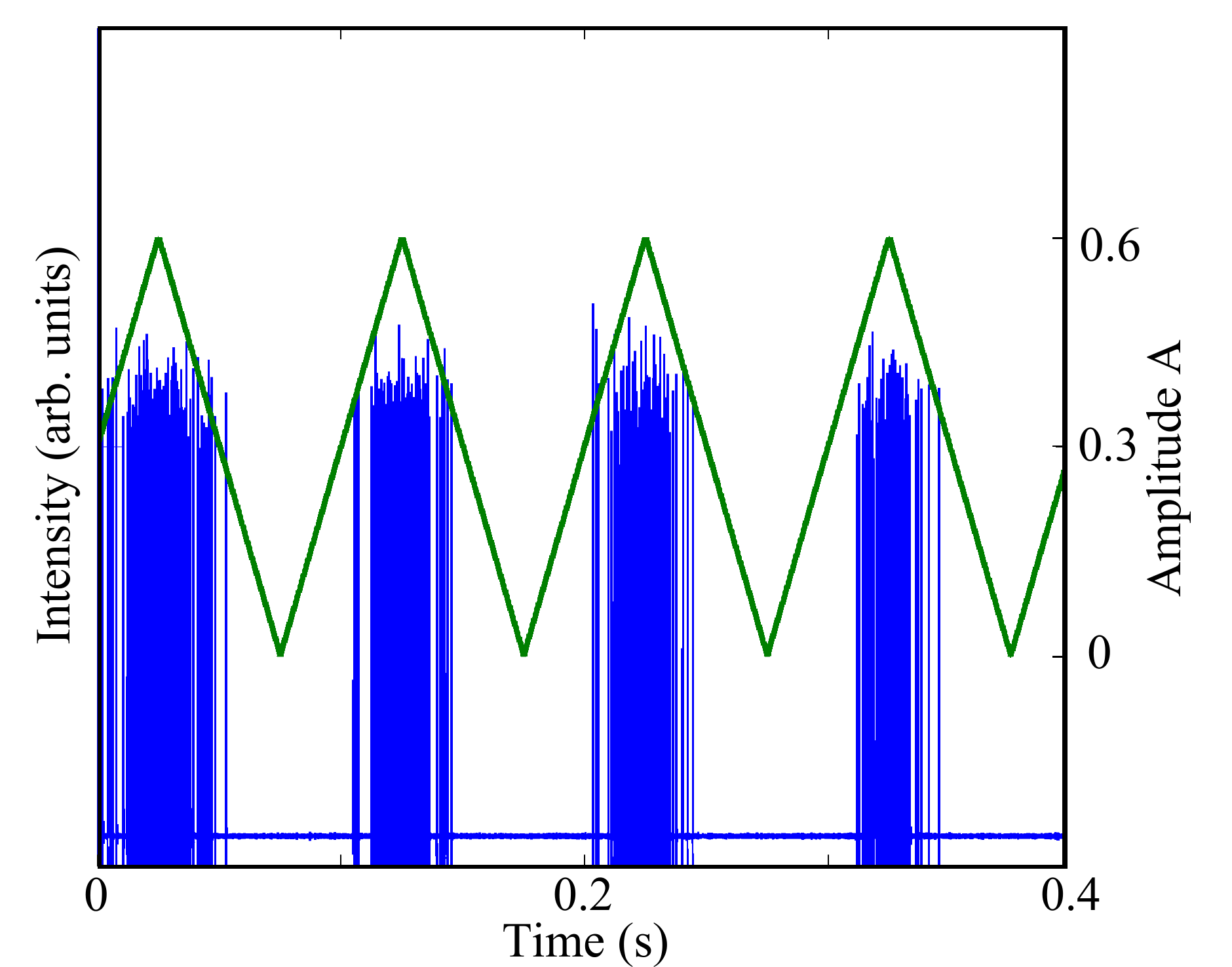}
\caption{Experimental beat-note intensity as a function of the amplitude $A$ of a kick to the detuning parameter.}
\label{fig:seuil}
\end{figure}
We have tested experimentally the excitable-like character of the observed dynamics, by applying abrupt, deterministic perturbations to the driving frequency of the AOM, which amounts to switching the detuning between two different values as in Fig.~\ref{fig:exca} (a). For Fig.~\ref{fig:seuil}, the system was prepared in the quiescent state, close to the bifurcation point, and then perturbations of increasing amplitude $A$ were applied. The response is plotted as a function of $A$. When $A$ is smaller than a certain value $A_{th}$, there is no response. When $A$ is sufficiently large, pulses of similar amplitude are emitted, irrespective of the precise value of $A$. The existence of a threshold value for $A$, and the all-or-none character of the response appear in Fig.~\ref{fig:seuil}. It can also be seen that $A_{th}$ varies from a ramp to the other. We attribute this to the fact that $A_{th}$ is determined by the distance to the bifurcation point i.e. by the value of $\Delta$, and in the experiment this parameter has inevitably some fluctuations, as explained in the discussion of Fig.~\ref{fig:Fig3}. 
\begin{figure}[htbp]
\centering
\includegraphics[width=\linewidth]{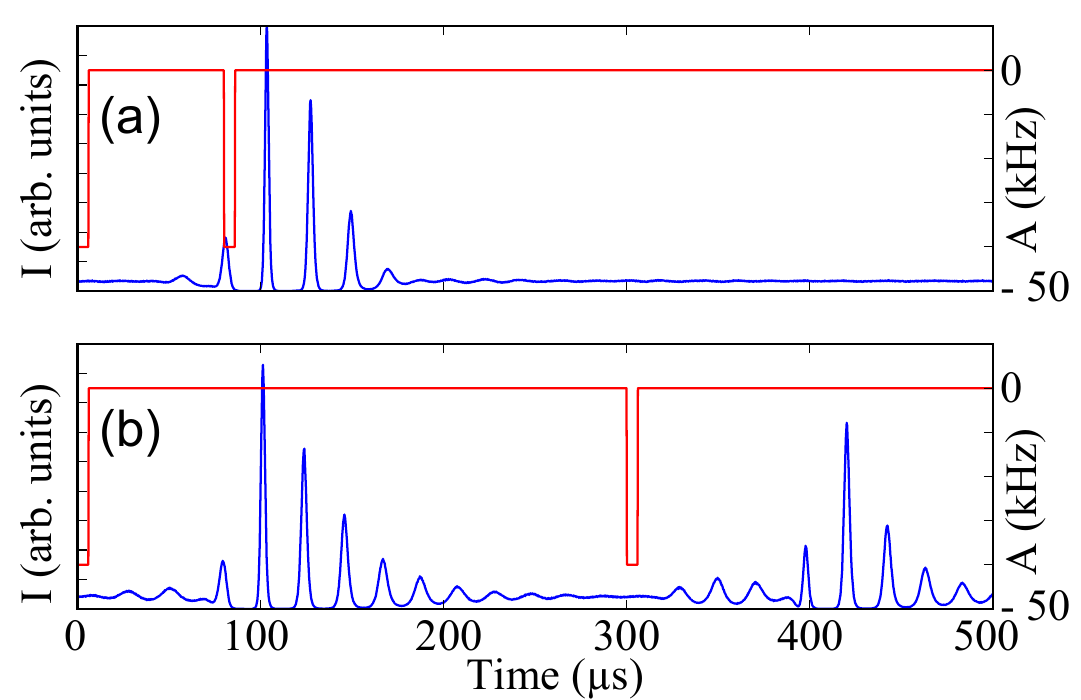}
\caption{Experimental response to two perturbations of the detuning parameter $\Delta$, separated by a different time interval.}
\label{fig:refract}
\end{figure}
We have also investigated the issue of the response to two consecutives excitations, to see if a property analogous to the refractory time of excitable systems could be observed.  It was not possible to obtain a meaningful experimental curve to be compared to the calculated one in Fig.~\ref{fig:excb} (b). Again, the reason is that the technical fluctuations of $\Delta$ have an important impact on the value of the absolute refractory time, expected from the theory to depend on the precise value of the distance from the bifurcation point. However, we were at least able to verify that, when the response to the first perturbation happens to develop when, or just before, the second kick is applied, then there is systematically no response to it (Fig.~\ref{fig:refract} (a)). On the contrary for sufficient delay there are two nearly identical responses. This is coherent with the interpretation  discussed in section~\ref{sec:theo}, and suggests that the proposed analogy with excitable systems is meaningful.  
\section{Conclusions}\label{sec:con}
In conclusion, we have provided experimental and numerical evidence of a behavior having several unusual features, and some properties that are reminiscent of excitable systems. We have observed large intensity spikes in the output of a driven opto-RF oscillator, and shown by measuring the phase variations that these events occur in the bounded-phase regime. We have interpreted our observations as noise-induced chaotic pulses. Numerical calculations indicate that, despite the presence of a chaotic attractor, the self-pulsating state associated to it is sufficiently regular to produce a response with properties resembling to those of excitable systems. The experimental results are coherent with the numerical findings, and suggests that the above picture is meaningful. 
These results illustrate the \textit{robustness} of excitable-like properties as a generic feature of nonlinear systems, since they can appear, as in our case, in presence of higher dimensionality, and even chaos.

\end{document}